\documentclass[aps, pra,twocolumn,groupedaddress]{revtex4-2}
\usepackage{CJK}
\usepackage{color}
\usepackage{graphicx}
\usepackage{dcolumn}
\usepackage{subfigure}
\usepackage{graphics}
\usepackage{amsmath}
\usepackage{amsfonts}
\usepackage{amssymb}
\usepackage{epsf}
\usepackage{bm}

\begin{document}
\begin{CJK*}{GB}{ } % Use default fonts from CJK (see below)

\title{Characterizing the coherence of three-atom entangled Rydberg states near F\"orster resonance}
\author{Dominick Frost, Calista Pearlstone, Anyssa El Mha\"ia, Lillian Gehres, Lucy Shamel, and Aaron Reinhard$^{*}$}

\address{
Department of Physics, Kenyon College, 201 North College Rd., Gambier, Ohio 43022, USA \\
$^*$Corresponding author: reinhard1@kenyon.edu\\
}

\date{\today}
\begin{abstract}
Neutral atoms are widely used to create scalable quantum computers and simulators. Such schemes rely on pairwise interactions among highly-excited, or Rydberg, atoms.  However, new applications may require genuine three-body interactions.  We have demonstrated a technique to isolate three-body interactions, leading to excitation of tripartite entangled Rydberg states.  We laser excite these states in a bulk gas using Ramsey interferometry and spin echo pulse sequences, and measure coherence by tracking the decay of the resulting fringes.  By varying principal quantum number near F\"orster resonance, we identify dipole-dipole interactions with a background of isolated Rydberg atoms as a source of inhomogeneous dephasing that cannot easily be ``echoed out.''
\end{abstract}

\maketitle		
\end{CJK*}

\section{INTRODUCTION}
Neutral atoms have become a leading platform for scalable quantum computation~\cite{Browaeys,Kaufman,Graham,Bluvstein} and simulation~\cite{Bernien,Semeghini,Scholl}.  The most significant progress has been achieved using optical tweezers.  Recent examples include mid-circuit error correction~\cite{Bluvstein,Singh}, annealing for optimization problems~\cite{Goswami,deOliveira}, simulations of quantum magnetism~\cite{Qiao}, and full tomography of highly nontrivial states~\cite{Bluvstein}.  Publicly-accessible arrays are even available for remote control~\cite{Dag}.   Alternative approaches involve encoding qubits into collective states of ensembles~\cite{Spong,Li,Shao,Jiao}.  While the approaches are distinct, all these applications rely on one core principle - binary interactions between highly excited, or Rydberg, atoms~\cite{Saffman,Reinhard}.

There is a growing sense in the field that genuinely three-body interactions might lead to new applications~\cite{Samajdar}.  These may involve multiqubit quantum gates~\cite{Tretyakov,Beterov,Ryabtsev,Ashkarin,Ashkarin2} or quantum simulation of new models in condensed matter~\cite{Sen,Nielsen} and high-energy physics~\cite{Hafezi}.  For example, controlled multiqubit operations may significantly reduce the number of gates needed to implement nontrivial algorithms~\cite{Dlaska,Tang}.  Recent works have proposed that, under carefully-controlled conditions, three-body Rydberg interactions can dominate over pairwise interactions~\cite{Samajdar,Faoro,Liu}.  However, examples are rare and coherence has been elusive.  In this paper, we demonstrate a novel three-body system.  Multi-atom dipole-dipole energy exchange can occur during Rydberg excitation using a narrowband laser~\cite{Reinhard2,Younge,Kondo}.  We show that, under appropriate conditions, one can coherently excite entangled superpositions of three atoms in distinct Rydberg states~\cite{Pohl,Eder,Yoda}.

We excite atoms from a ground state, $\vert g \rangle$, to a target state Rydberg state, $\vert d \rangle$, via a resonant optical pulse.  Two-particle states, $\vert dd \rangle$, are coupled via dipole-dipole interactions to product states, $\vert pf \rangle$ and $\vert fp \rangle$.  The labels $d$, $p$, and $f$ indicate the angular momentum quantum numbers of the single-atom states. The strongest coupling is to states with principal quantum numbers differing by 1 or 2.  A F\"orster resonance occurs when one coupled two-atom state has a small energy defect: $\Delta E =2 \times E_{d} - (E_{p}+E_{f}) \approx 0$.  Here, the atom-atom interactions are enhanced, and the two-body eigenstates are mixtures of bare atomic states. In rubidium, such a resonance occurs near $n=43$, where $\vert d \rangle = nD_{5/2}$, $\vert p \rangle = (n+2)P_{3/2}$, and $\vert f \rangle = (n-2)F_{7/2}$~\cite{Reinhard,Walker}.

Nontrivial energy-exchange occurs near F\"orster resonance. If the two-atom energy shift is smaller than the laser linewidth, the dipole blockade breaks down and one can excite pairs of atoms to an entangled state $\vert M_2 \rangle = a_1 \vert dd \rangle + a_2 \vert pf \rangle + a_3 \vert fp \rangle$, where $a_i$ are probability amplitudes ~\cite{Reinhard2,Younge,Kondo,Eder}.  However, a single optical pulse can also drive transitions between the three-atom ground state $\vert ggg \rangle$ and a triply-excited state, $\vert M_3 \rangle$, via electrostatic couplings to virtual levels.  This entangled state has zero energy shift, and has the form $\vert M_3 \rangle = c_1 \vert dpf \rangle + c_2 \vert dfp \rangle + c_3 \vert pdf \rangle + c_4 \vert pfd \rangle + c_5 \vert fdp \rangle + c_6 \vert fpd \rangle$, where $c_i$ are probability amplitudes~\cite{Pohl}.  Given appropriate values for the energy defect, pulse duration, and Rabi frequency, $\vert M_3 \rangle$ can be excited with relatively high probability~\cite{Yoda}.

Coherence in $\vert M_3 \rangle$ has been detected using an optical rotary echo technique~\cite{Yoda}.  In the present work, we directly observe the coherence of $\vert M_3 \rangle$ in multiple observables, and we fully characterize the dominant sources of decoherence.  We use Ramsey interferometry to measure the coherence time and spin echo to quantify the strength of inhomogeneous dephasing.  We further show that dipole-dipole interactions with a background of single Rydberg atoms are a source of inhomogeneous dephasing that cannot be ``echoed out.''  To our knowledge, no other system has demonstrated coherent dipole-dipole energy exchange among Rydberg atoms beyond the two-atom level.  Thus, our work might find use in quantum information and simulation applications requiring multi-body physics. Because the coherence is clear even in a disordered bulk gas, it may be especially useful in situations where single-atom control is difficult or impractical.

\section{EXPERIMENTAL PROCEDURE}

We collect ultracold $^{85}$Rb atoms in a far off-resonant optical dipole trap.  We achieve ground-state atom densities of $\rho \sim 5 \times 10^{10}$~cm$^{-3}$ by turning off the dipole trap beam and allowing the atoms to freely expand for a few tens of $\mu$s before they are excited to Rydberg states~\cite{Younge, Eder, Yoda}.  The atoms in the dipole trap have temperature $T \sim 100~\mu$K, and our pulse sequences take $\leq 2~\mu$s.  Therefore, the atoms are effectively frozen in space during Rydberg excitation.  The cloud has rms widths of about $9~\mu$m in the horizontal directions and $250~\mu$m in the vertical direction.  We excite atoms via a two-step optical transition $5S_{1/2} \rightarrow 6P_{3/2} \rightarrow nD_{5/2}$, with intermediate state detuning $\Delta=50$~MHz.  Therefore, pairs of coincident laser pulses drive an effective one-photon transition from $5S_{1/2} \rightarrow nD_{5/2}$. The lower-transition beam has $\sigma +$ polarization and the upper-transition has $\pi$ polarization. The upper-transition laser is frequency stabilized to a fixed-length Fabry P\'erot interferometer whose frequency resonances are tuned by changing the pressure of the gas inside~\cite{Orr}.  The beams are perpendicular to each other and to the long axis of the cloud.  Pulses are created by amplitude modulating acousto-optic modulators (AOMs) in each beam.

In Ref.~\cite{Yoda}, we observed coherent excitation of $\vert M_3 \rangle$ with probability $\sim 15\%$.   The excitation probability is a sensitive function of atom-atom separation, so the excitation pulses ``picked out'' atoms separated by a certain distance.  This allowed us to see one cycle of Rabi oscillation in a disordered sample.  In the present work, for an effective Rabi frequency of 0.5~MHz on $\vert g \rangle \rightarrow \vert d \rangle$, we achieve an effective $\pi$ pulse on the transition $\vert ggg \rangle \rightarrow \vert M_3 \rangle$ in approximately 500~ns.  We use two different pulse sequences to coherently drive $\vert M_3 \rangle$ on the Bloch sphere. The first sequence is Ramsey interferometry, where two (effective) $\pi/2$ pulses are applied with a time separation of $\tau$.  The second sequence is spin echo, where we add an (effective) $\pi$ pulse halfway between the two Ramsey pulses.  For both pulse sequences, the last $\pi/2$ pulse has an adjustable relative phase, $\phi$.  To achieve this phase difference, we phase modulate the RF signal driving the AOM in the upper-transition beam path.

To distinguish between resonant energy exchange among pairs (producing $\vert M_2 \rangle$) and triples (producing $\vert M_3 \rangle$), we use a method that relies on shot-to-shot fluctuation in the number of energy exchange events during Rydberg excitation~\cite{Eder,Yoda}.  We excite atoms to an $nD_{5/2}$ Rydberg state using laser pulses tuned to two-photon resonance.  During excitation, atoms may exchange energy via dipole-dipole interactions and end up in $\vert M_2 \rangle$ or $\vert M_3 \rangle$.  We measure the state of each individual Rydberg atom immediately after excitation using state-selective field ionization (SSFI) spectroscopy~\cite{Ducas,Gallagher}.  A pulse counter records the number of excitations in each of two independent counting gates, for each of 1001 instances of the experiment.  The ``P Gate'' counts atoms in the $(n+2)P_{3/2}$ product state, or $\vert p \rangle$, while the ``T Gate'' counts the total number of Rydberg atoms.  From this data, we build a 2D histogram (``sorted graph'') of the total number of excitations as a function of the number in $\vert p \rangle$.  We fit each sorted graph to a straight line.  The slope of this line indicates the number of additional Rydberg excitations created each time another atom is detected in $\vert p \rangle$.  A small value of slope indicates that energy exchange is dominated a two-body process ($\vert M_2 \rangle$), since two additional Rydberg atoms (in $\vert p \rangle$ and $\vert f \rangle$) are created for each atom in $\vert p \rangle$.  A large value of slope indicates the presence of three-atom entangled states, $\vert M_3 \rangle$, because three additional Rydberg atoms (one each in $\vert d \rangle$, $\vert p \rangle$, and $\vert f \rangle$) are created for each atom in $\vert p \rangle$.

\begin{figure}[htp]
\centerline{ \scalebox{0.1375} {\includegraphics{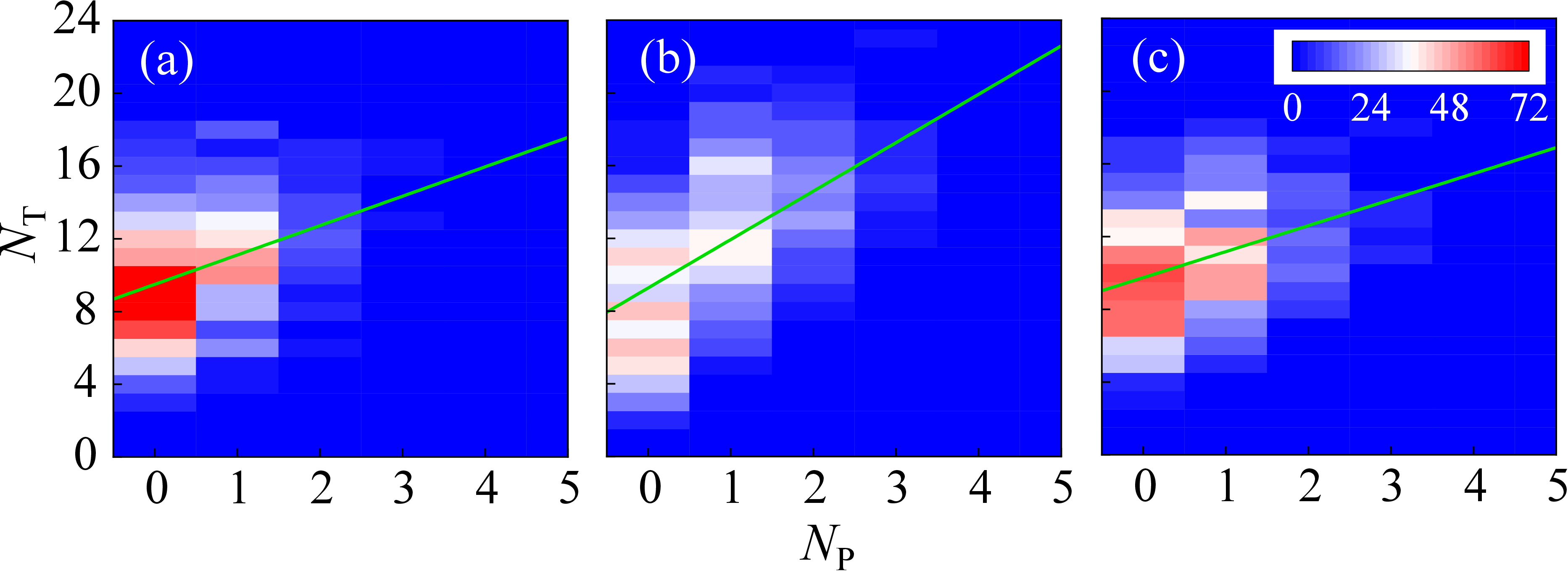}}} \caption{ \label{Fig2} Sorted graphs for excitation to $42D_{5/2}$ using a Ramsey sequence with 20~ns pulse separation. These graphs show the total number of excitations, $N_T$, on the y axis and the number in product state $\vert p \rangle$, $N_P$, on the x-axis.  The false color indicates how many of each $\{N_P, N_T\}$ were detected (scale in upper right).  The green line is a fit, from which we extract the slope. The relative phase of the second Ramsey pulse is $-\pi$ in (a), 0 in (b), and $\pi$ in (c).}
\end{figure}

Example sorted graphs are shown in Fig.~\ref{Fig2}.  The excitation lasers are tuned to resonance with $42D_{5/2}$, using a Ramsey sequence with $\tau=20$~ns.  The false color scale indicates the number times we measure $N_P$ atoms in the P Gate and $N_T$ atoms in the T Gate.  The green lines are a least-squares linear fit to the data.  One can see that as the relative phase of the second Ramsey pulse is varied from $-\pi$ to $\pi$, the slope of the sorted graphs go from low to high and then back to low.  This suggests that we might see a Ramsey fringe in the slope as a function of the phase of the second Ramsey pulse.  This would indicate that the mechanism causing energy exchange near F\"orster resonance (two-body vs three-body) is highly sensitive to the phase of the second pulse. While general trends are clear, the precise (non-integer) values of slope are predicted by a Monte Carlo model (see below).

\section{RESULTS AND DISCUSSION}

\begin{figure}[htp]
\centerline{ \scalebox{0.41} {\includegraphics{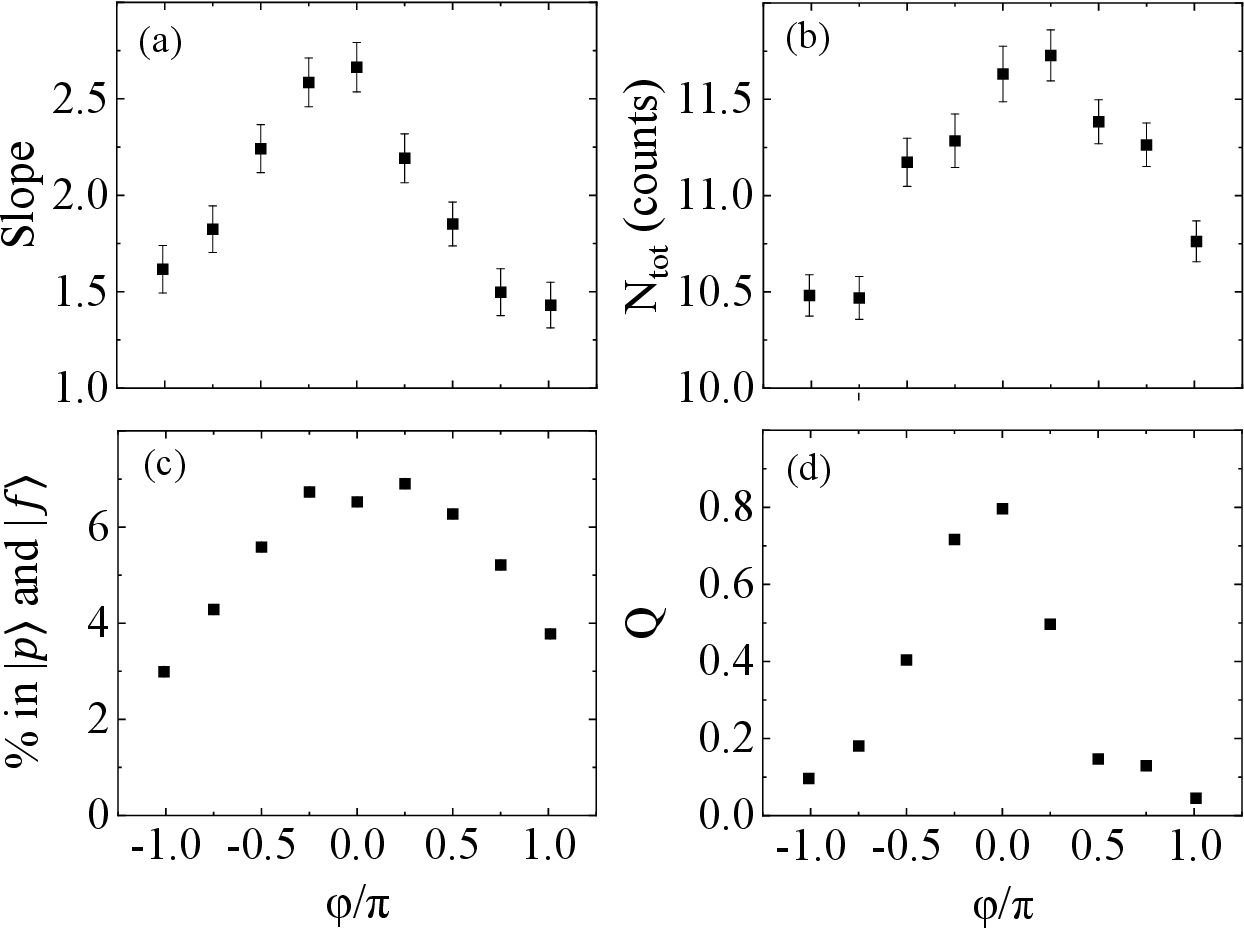}}} \caption{ \label{Fig3} Slopes of the sorted graphs (a), total number of Rydberg counts (b), mixing fraction (c), and Mandel $Q$ (d) as a function of the phase of the second Ramsey pulse, for excitation to the $42D_{5/2}$ state with $\tau=20$~ns. Error bars are the one-sigma uncertainty resulting from a least-squares fit to the raw data.}
\end{figure}

To confirm the presence of Ramsey fringes, we plot the slopes of the sorted graphs as a function of $\phi$ for $\tau=20$~ns in Fig.~\ref{Fig3}a.  Since the two pulses are separated by a time much less than their durations, these conditions are similar to a rotary echo~\cite{Raitzsch,Younge2,Yoda}.  With a relative phase of zero the slope of the sorted graphs is maximum, since two (nearly) sequential, in-phase $\pi/2$ pulses create a $\pi$ pulse.  This drives atoms from $\vert ggg \rangle \rightarrow \vert M_3 \rangle$ with maximum probability.  As $\vert \phi \vert$ increases from zero, the second pulse partially reverses the excitation produced by the first pulse.  For $\vert \phi \vert = \pi$ the second pulse entirely undoes the excitation.  This leads to the fringe characteristic of a short separation Ramsey (or rotary echo) experiment~\cite{Yoda}.

To provide further evidence for coherence, we present panels b-d of Fig.~\ref{Fig3}.  In panel b, we plot the total number of Rydberg excitations. In panel c, we plot the mixing fraction,  defined as the fraction of all Rydberg excitations detected in $\vert p \rangle$ and $\vert f \rangle$ states~\cite{note}. It indicates how many Rydberg excitations mix into product states due to dipole-dipole energy exchange. Finally, in panel d we plot the Mandel $Q$ parameter, which is defined as $Q=\sigma^2 / \bar{N_T}-1$, where $\sigma^2$ is the dispersion and $\bar{N_T}$ is the mean number of total excitations.  It represents the width of the excitation number distribution. In every case, we see the same characteristic Ramsey fringe.  This can be understood as follows. Because our cloud is much larger than the blockade volume, optical excitation may result in (i) single atoms being excited to the target state, $\vert d \rangle$, (one-body process),  (ii) close pairs being excited to $\vert M_2 \rangle$ at zero crossings in the pair potentials (two-body process)~\cite{Eder}, or (iii) triples being excited to $\vert M_3 \rangle$ (three-body process)~\cite{Yoda}.  By varying the phase of the second Ramsey pulse ($\phi$) we control the likelihood of exciting $\vert M_3 \rangle$.  When $\vert \phi \vert \approx \pi$, excitation to $\vert M_3 \rangle$ is coherently reversed by the second Ramsey pulse.  Therefore, the number of excitations, fraction of atoms in $\vert p \rangle$ and $\vert f \rangle$, and level of fluctuations are characteristic of the values they would have with no excitation to three-body states.  However, as $\vert \phi \vert \rightarrow 0$ the probability to excite $\vert M_3 \rangle$ approaches its maximum.  As this happens, all three of the observables in panels b-d go to a higher value.  The total number of excitations increases as the probability to add some three-atom excited states is increased. The mixing fraction increases when we go from only two-body energy exchange processes (for closely separated pairs) to the possibility of additional, three-body energy exchange events. Finally, the Mandel $Q$ increases because there are more fluctuations in total atom number when all three processes (one-, two-, and three-body) are present.

We see additional evidence for excitation of $\vert M_3 \rangle$ if we demand internal consistency between the panels in Fig.~\ref{Fig3}.  In panel c, there is a difference of 4\% in the fraction measured in $\vert p \rangle$ and $\vert f \rangle$ as $\phi$ is varied.  Given the values in panel b, this would predict an amplitude of about 0.42 counts in $N_{\rm{tot}}$ vs $\phi$ if we assume the change is due to two-body interactions.  If only $\vert p \rangle$ and $\vert f \rangle$ atoms are added together, we would expect one additional excitation in 4\% of events, compared with excitation of only $\vert d \rangle$.  However, if we assume three-body interactions, the data in panel b would predict an amplitude of about 0.84. If one each of $\vert d \rangle$, $\vert p \rangle$, and $\vert f \rangle$ are added together, we would expect two additional excitations in 4\% of events, compared with excitation of only $\vert d \rangle$. The measured amplitude in panel b is 1.1. The small discrepancy might be explained by reversal of single-atom excitation to $\vert d \rangle$ with varying $\phi$~\cite{Younge}.  Thus, panels b and c are consistent only if we assume changes in product state population arise predominantly due to three-body interactions.

\begin{figure}[htp]
\centerline{ \scalebox{0.475} {\includegraphics{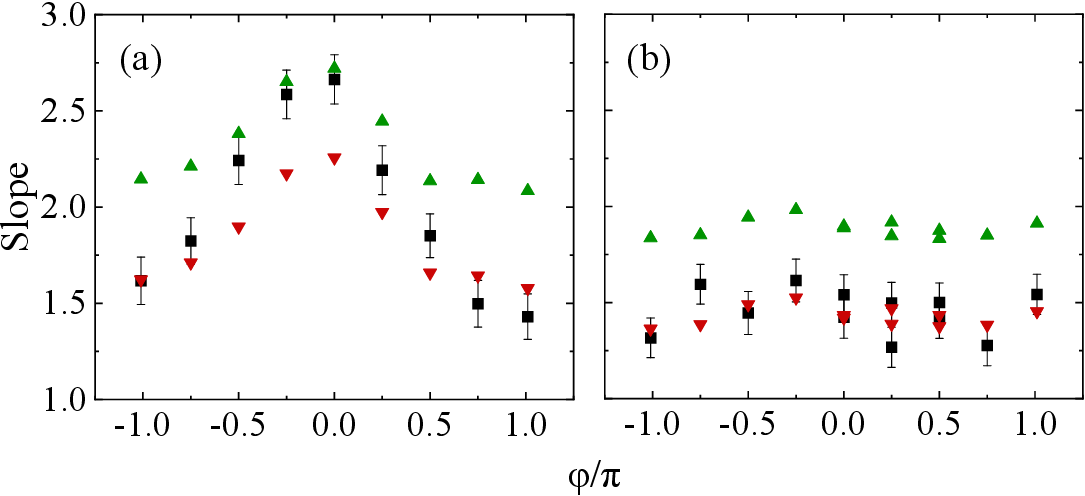}}} \caption{ \label{Fig4} Slopes of the sorted graphs for excitation to the $42D_{5/2}$ state with $\tau=20$~ns (a) and 600~ns (b). The black squares are data, the red down triangles are the result of a Monte Carlo simulation assuming two-body energy exchange, and the green up triangles are the result of a Monte Carlo simulation assuming three-body energy exchange.}
\end{figure}

The obvious next step is to change the time between pulses, $\tau$, to measure coherence time.  We could, in principle, consider Ramsey fringes in any one of the observables in Fig.~\ref{Fig3}.  In practice, we focus on the slope of the sorted graphs, as this quantity is least sensitive to day-to-day changes in experimental conditions.  For the other observables, small changes in excitation laser intensity, beam overlaps, or ground state atom density can make it difficult to compare data taken on different days. In Fig.~\ref{Fig4} we plot the slope of the sorted graphs as a function of $\phi$ for $\tau=20$~ns (a) and for $\tau=600$~ns (b).  It is immediately apparent that the strong fringe present for $\tau=20$~ns is gone for $\tau=600$~ns.  Apparently, the coherence time of $\vert M_3 \rangle$ is much less than 600~ns.

It is not straightforward to directly interpret the slopes of the sorted graphs.  In particular, why is the value not an integer? The reason is that non-unity detector efficiency causes the measured slope to depend on Mandel $Q$. If one shot of the experiment produces $N_P$ atoms in the P-gate and $N_T$ atoms in the T-gate, fewer than these number will actually be measured.  Therefore, in the sorted graph, the pair $\{N_P, N_T\}$ will be measured closer to the origin than it would if detector efficiency were 1.  However, one cannot detect fewer than zero counts in either gate.  The result is that higher Mandel $Q$ is correlated with a higher slope.

To make this quantitative, we simulate excitation, energy exchange, and detection using a Monte Carlo model.  Each iteration of the simulation draws a random number of excitation events from a Gaussian distribution.  The width of this distribution is chosen to reproduce the measured Mandel $Q$ parameter after (simulated) detection.  We choose a distribution that yields a mean of about 12 detected excitations, to match experiment.  However, our results are highly insensitive to this value.  We perform two types of simulations: one with two-body dominated energy exchange and one with three-body dominated energy exchange. For a two-body simulation, we randomly assign each excitation event to either a single atom in $\vert d \rangle$ or to two atoms, one each in $\vert p \rangle$ and $\vert f \rangle$ (due to excitation of $\vert M_2 \rangle$).  For the three-body simulation, we randomly assign each excitation event to either a single atom in $\vert d \rangle$ or to three atoms, one each in $\vert d \rangle$, $\vert p \rangle$, and $\vert f \rangle$ (due to excitation of $\vert M_3 \rangle$).  The probability to excite atoms to $\vert d \rangle$, $\vert p \rangle$, and $\vert f \rangle$ are given by the experimentally-measured mixing fraction.  After every excitation event, the program decides if each Rydberg count is recorded, using our microchannel plate's detector efficiency~\cite{Wiza}.

For each Monte Carlo point in Fig.~\ref{Fig4}, we run the simulation $5 \times 10^4$ times, and plot the total number of Rydberg excitations detected as a function of the number in $\vert p \rangle$.  We fit this simulated sorted graph to a line and extract the slope, just as in the experiment.  Therefore, the Monte Carlo model predicts a value of slope, given the assumption of a dominant two-atom or three-atom model for quantum state mixing, and the experimentally measured mixing fraction and Mandel $Q$.  Note that there are no free parameters.

As seen in Fig.~\ref{Fig4}a, the slopes of the sorted graphs are best described by dominant two-body energy exchange for $\vert \phi \vert \approx \pi$ and they transition to three-body energy exchange as $\vert \phi \vert \rightarrow 0$.  We attribute the change to an increasing probability to excite $\vert M_3 \rangle$, as shown by direct calculation in Ref.~\cite{Yoda}.  The behavior is different for large $\tau$, as seen in Fig.~\ref{Fig4}b.  In this case, both the data and simulated points are a constant function of $\phi$.  There are no Ramsey fringes, as coherence has been lost. However, the value of the slope is well-explained by two-body energy exchange.  This adds weight to the argument that we can excite the state $\vert M_3 \rangle$ coherently within a window of order hundreds of ns.  Outside of this window, population can only be transferred to $\vert p \rangle$ and $\vert f \rangle$ via incoherent excitation of $\vert M_2 \rangle$ at short-distance zero crossings in the Rydberg pair potentials~\cite{Eder}.

In each panel of Fig.~\ref{Fig4} the two-body and three-body simulations have the same qualitative shape.  This is because of the dependence of the slope of the sorted graph on Mandel $Q$.  As one can see in Fig.~\ref{Fig3}, the Mandel $Q$ has a relative maximum, following the probability to excite $\vert M_3 \rangle$.  As Mandel $Q$ is an input to our simulation, the predicted slopes will have the same shape.  Using this three-body-produced Mandel $Q$ to simulate two-body energy exchange thus
results in an erroneously high prediction for the two-body slope. However, even with this increase, the two-body prediction undershoots the data points, illustrating the dominance of the three-body process for small relative phases.

\begin{figure}[htp]
\centerline{ \scalebox{0.60} {\includegraphics{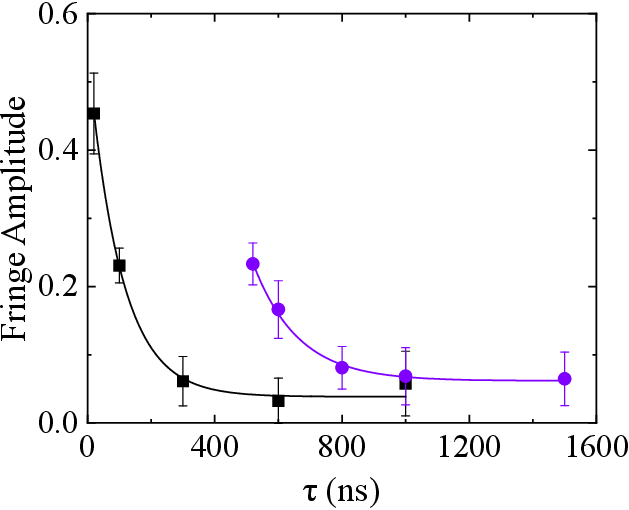}}} \caption{ \label{Fig5} Amplitude of Ramsey fringes in slope for excitation to $41D_{5/2}$ as a function of pulse separation, $\tau$.  The black squares are for a Ramsey sequence and the purple circles are for a spin-echo sequence.}
\end{figure}

To quantify the coherence time of $\vert M_3 \rangle$, we plot the amplitude of the Ramsey fringes in slope vs $\phi$ as a function of the time between Ramsey pulses, $\tau$.  This is shown in Fig.~\ref{Fig5} as black squares for excitation to $41D_{5/2}$.  A least-squares exponential fit indicates a coherence time of $T2^{*} = 100 \pm 10$~ns, consistent with our earlier estimate that the coherence would last significantly less than 600~ns.  The coherence is robust and clearly measurable in the ``messy'' environment of a disordered bulk gas.  However, given that our $\pi/2$ pulses are 250~ns long, this is a relatively short coherence time.

We next explore the factors that limit the coherence. As a first step, we performed a spin echo sequence.  Here, we apply a single $\pi$ pulse, halfway between the two Ramsey pulses, which are separated in time by $\tau$.  The $\pi$ pulse reverses inhomogeneous dephasing by reflecting the dephased Bloch vectors about the x-axis and reversing the direction of their free evolution.  At the ``echo time,'' or the time of the second Ramsey pulse, all Boch vectors will momentarily re-phase and a rotation can be executed coherently.  In Fig.~\ref{Fig5} we plot the amplitude of the Ramsey fringes in slope as a function of $\tau$ (purple circles) for excitation to $41D_{5/2}$.  It is immediately apparent that the coherence time is longer.  A least squares exponential fit yields a coherence time of $T2 = 140 \pm 20$~ns.  For $\vert d \rangle = 41D_{5/2}$, inhomogeneous dephasing reduces the coherence time by 30\%.

We compare this value of $T2$ to the limit imposed by the frequency stability of our laser sources.  The lower-transition laser is current-dithered and peak-locked to a $5S_{1/2} \rightarrow 6P_{3/2}$ saturated absorption feature in a heated vapor cell.  The upper-transition laser is current-dithered and peak-locked to a transmission resonance in our Fabry P\'erot cavity, which has finesse of about 130~\cite{Orr}.  Our effective excitation linewidth is 1.3~MHz, dominated by the frequency stability of the lower-transition laser.  If laser frequency stabilization were the limiting factor, we would have $T2=\frac{1}{\pi \Delta \nu}$, where $\Delta \nu$ is the laser linewidth.  Therefore we would expect a coherence time of about 245~ns.  Clearly another factor is limiting the coherence.

\begin{figure}[htp]
\centerline{ \scalebox{0.415} {\includegraphics{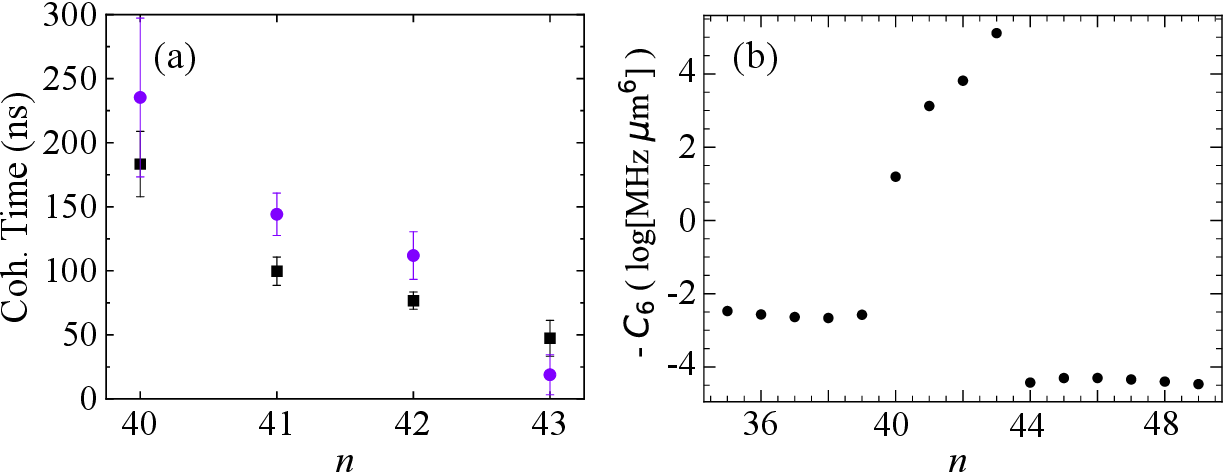}}} \caption{ \label{Fig6} (a) Coherence time as a function of the principal quantum number, $n$, of the target state.  The black squares represent $T2^*$ and the purple circles represent $T2$.  (b) $-C_6$ as a function of $n$ for pairwise dipole-dipole interaction between two $nD_{5/2}$, $m_{\rm j} = 5/2$ Rydberg atoms positioned along the shared quantization axis.}
\end{figure}

One possibility is dipole-dipole interactions between $\vert M_3 \rangle$ and the background of atoms excited to $\vert d \rangle$.  $\vert M_3 \rangle$ is an entangled superposition of terms of the form $\vert dpf \rangle$ or $\vert pfd \rangle$, where $\vert d \rangle$ is the target state and $\vert p \rangle$ and $\vert f \rangle$ are the product states.  Each atom in the triple interacts with background atoms (in $\vert d \rangle$) via either van der Waals or resonant dipole-dipole interactions.  However, the background is randomly distributed in space.  Atom-atom interactions are not reversed by a spin-echo sequence, as each interaction partner has its phase evolution reversed. Therefore, inhomogeneous interactions with the background might be a limiting factor.  To investigate this effect, we plot the coherence times, $T2^{*}$ and $T2$ as a function of the principal quantum number of the target state, $\vert d \rangle$, in Fig.~\ref{Fig6}a.  Since we excite atoms to target $nD_{5/2}$ states near F\"orster resonance, the strength of dipole-dipole interactions are a strongly-varying function of $n$.  The interactions are strongest at $n=43$ and decrease as $n$ is decreased until, at $n=40$, the two dominant interaction channels nearly cancel and the interaction strength is low~\cite{Reinhard}.  This is summarized in Fig.~\ref{Fig6}b, where we plot $-C_6$ as a function of $n$ for pairs of atoms in $nD_{5/2}$.  Since the van der Waals interaction energy for two atoms separated by $R$ is $E_{\rm{VdW}}(R) = -C_6/R^6$, $-C_6$ is proportional to strength of the interatomic interactions.

We can see from Fig.~\ref{Fig6}a that $T2^{*} \approx T2$ and is very short at $n=43$, where the van der Waals interactions are strong.  As $n$ decreases, the interaction strength goes down by several orders of magnitude and the coherence times grow.  For $n=41$ and 42, the coherence times are of order 100-150~ns, and spin echo produces $T2>T2^{*}$.  At $n=40$, the two times are not statistically different, owing mostly to the large error bars on $T2$.  Because the coherence times are about a factor of 4 shorter than the total time the pulses are on, the data are somewhat noisy.

There are two main takeaways.  First, the coherence time is a strongly varying function of $n$ near F\"orster resonance, indicating that dipole-dipole interactions are a limiting factor.  Second, at $n=40$ where dipole-dipole interactions are weak, $T2 = 235 \pm 60$~ns, consistent with the upper limit predicted by our modest laser frequency control (245~ns).  Thus, one might expect longer coherence times with two improvements: better laser frequency stabilization and removal of dipole-dipole interactions with the inhomogeneous background.  The latter could be achieved in more than one way, such as with an optical tweezer array that isolates only three atoms or by applying an engineered pulse sequence, such as WAHUHA, to ``echo-out'' the effect of dipole-dipole interactions in a bulk gas~\cite{Waugh,Duer,Zhou}.  WAHUHA is not available to us at present, as it requires six $\pi/2$ pulses, which would stretch the minimum Ramsey separation $\tau$ to about 6 times the current longest coherence time.

\section{CONCLUSION}

We have quantified the coherence times of three-atom entangled Rydberg states excited near F\"orster resonance.  We used Ramsey interferometry and spin echo to demonstrate coherence in several observables, each of which indicate the presence of the entangled three-atom states predicted in Ref.~\cite{Pohl}.  A Monte Carlo model confirmed that the mechanism causing energy exchange may be coherently controlled using the phase of the last laser pulse.  We determined the factors that limit coherence time and, more importantly, showed that there are good prospects for longer coherence with improved laser frequency stabilization and reduced background interactions. The latter goal may be achieved by using an optical tweezer array to isolate three single atoms, or by using tailored pulse sequences such as WAHUHA. Given our short coherence times, such multi-pulse sequences will only be possible by significantly reducing the $\pi$ pulse durations.  This might be achieved with significantly higher optical power (not available to us at present) or by moving the coherent manipulation into the Rydberg domain, where microwave transitions between closely-spaced levels allow for high Rabi frequencies.

{\em Acknowledgements.}  The authors acknowledge valuable input from Bryce Gadway and Karen Wadenpfuhl. This work was supported by NSF Grant PHY-2204899.


\begin{thebibliography}{99}
\bibitem{note} The mixing fraction is calculated as twice the integral of the $p$-peak in the SSFI spectrum, divided by the total integral.
\end{thebibliography}
\end{document}